\documentclass{webofc}

\usepackage[varg]{txfonts}   
\usepackage{hyperref}
\usepackage{url}
\usepackage{geometry} 
\hypersetup{colorlinks=true,citecolor=blue,urlcolor=blue,linkcolor=blue}
\usepackage{etoolbox} 

\makeatletter
\pretocmd{\maketitle}{\vspace*{-3.0cm}}{}{}
\makeatother
\begin{document}
\title{Search for heavy resonances decaying into two Higgs bosons in the $\mathrm{b}\bar{\mathrm{b}}\,\tau^{+}\tau^{-}$ final state in proton-proton collisions at $\sqrt{s}=13~\text{TeV}$ with the CMS detector}
%
%

\author{\firstname{Ganesh} \lastname{Parida on behalf of the CMS collaboration}\inst{1}\fnsep\thanks{\email{ganesh.parida@cern.ch}}}

\institute{University of Wisconsin-Madison}

\abstract{A search is presented for massive narrow-width resonances in the mass range of $1\text{-}4.5\,\text{TeV}$ decaying into pairs of Higgs bosons (HH), using proton-proton collision data at a center-of-mass energy of $13\,\text{TeV}$ collected with the CMS detector at the LHC during the $2016\text{-}2018$ data-taking. The data correspond to an integrated luminosity of $138~\mathrm{fb}^{-1}$. The analysis targets final states where one Higgs boson decays into a pair of bottom quarks and the other into a pair of tau leptons, $\mathrm{X}\rightarrow\mathrm{HH}\rightarrow \mathrm{b}\bar{\mathrm{b}}\,\tau^{+}\tau^{-}$. The observed data are found to be consistent with standard model background expectations. Upper limits at $95\%$ confidence level (CL) are set on the production cross section for resonant HH production for masses between $1$ and $4.5\,\text{TeV}$. This analysis sets the most sensitive LHC limits to date on $\mathrm{X}\rightarrow\mathrm{HH}\rightarrow \mathrm{b}\bar{\mathrm{b}}\,\tau^{+}\tau^{-}$ decays in the mass range of $1.4$ to $4.5\,\text{TeV}$.}
\maketitle
\section{Introduction}
\label{intro}
The discovery of the Higgs boson by the ATLAS~\cite{ATLAS:2013xga} and CMS~\cite{CMS:2012qbp} experiments established the standard model (SM) mechanism of electroweak symmetry breaking, yet several open questions motivate new physics at the TeV scale.
Many beyond the standard model (BSM) scenarios predict heavy resonances (X) decaying to Higgs boson pairs (HH), such as, warped extra-dimension models proposed by Randall and Sundrum~\cite{randall1999large}, both spin-0 scalars~\cite{goldberger1999modulus} and spin-2 Kaluza--Klein gravitons~\cite{davoudiasl2000phenomenology}. These processes, if they exist, can significantly enhance the HH production rate.

This analysis adopts a unified, model-independent strategy for both spin-0 and spin-2 resonances, targeting $\mathrm{X} \to \mathrm{HH} \to \mathrm{b\bar{b}}\,\mathrm{\tau\tau}$ in the mass range 1--4.5~TeV using the full Run~2 CMS dataset (2016--2018). Previous CMS searches in this final state, based on partial Run~2 data (35.9 $\mathrm{fb^{-1}}$)~\cite{2016cmsheavybbtautau}, observed no significant deviations from SM expectations. The corresponding full Run~2 ATLAS searches~\cite{aad2020reconstruction,Run2atlasheavybbtautau} also reported results consistent with the hypothesis of background-only.

\section{Signal topology and physics object reconstruction and identification}
\label{topology}

This analysis assumes a narrow resonance width, i.e., the natural width of the
resonance is significantly smaller than the experimental resolution. In the mass
range considered, the Higgs bosons are frequently produced with large transverse
momenta (Lorentz boosted regime). In the
$\text{H}\to \text{b}\bar{\text{b}}$ decay, the two bottom quarks are then emitted with a small angular
separation and, after hadronization, often merge into a single large-radius jet
($R=0.8$). These AK8 jets are identified with the \textsc{ParticleNet} jet
tagger~\cite{qu2020jet}. The mass regression
($M_{\mathrm{H(b\bar{b})}}$) branch of \textsc{ParticleNet} is also used in the
definition of the signal region (SR) and sidebands (SB), as described later in
Sec.~\ref{sec:bkg_estimation_results}.

\begin{figure}[tbp]
\centering
    \includegraphics[width=0.48\textwidth]{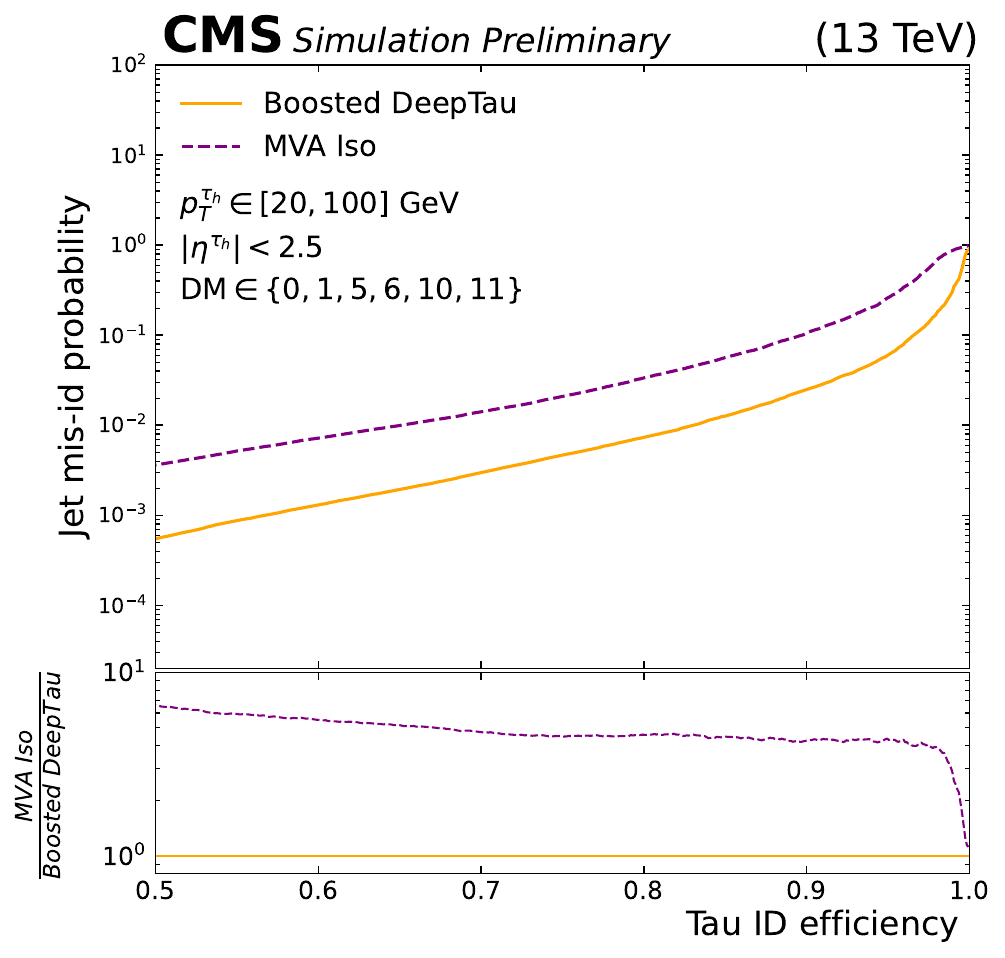}
    \includegraphics[width=0.48\textwidth]{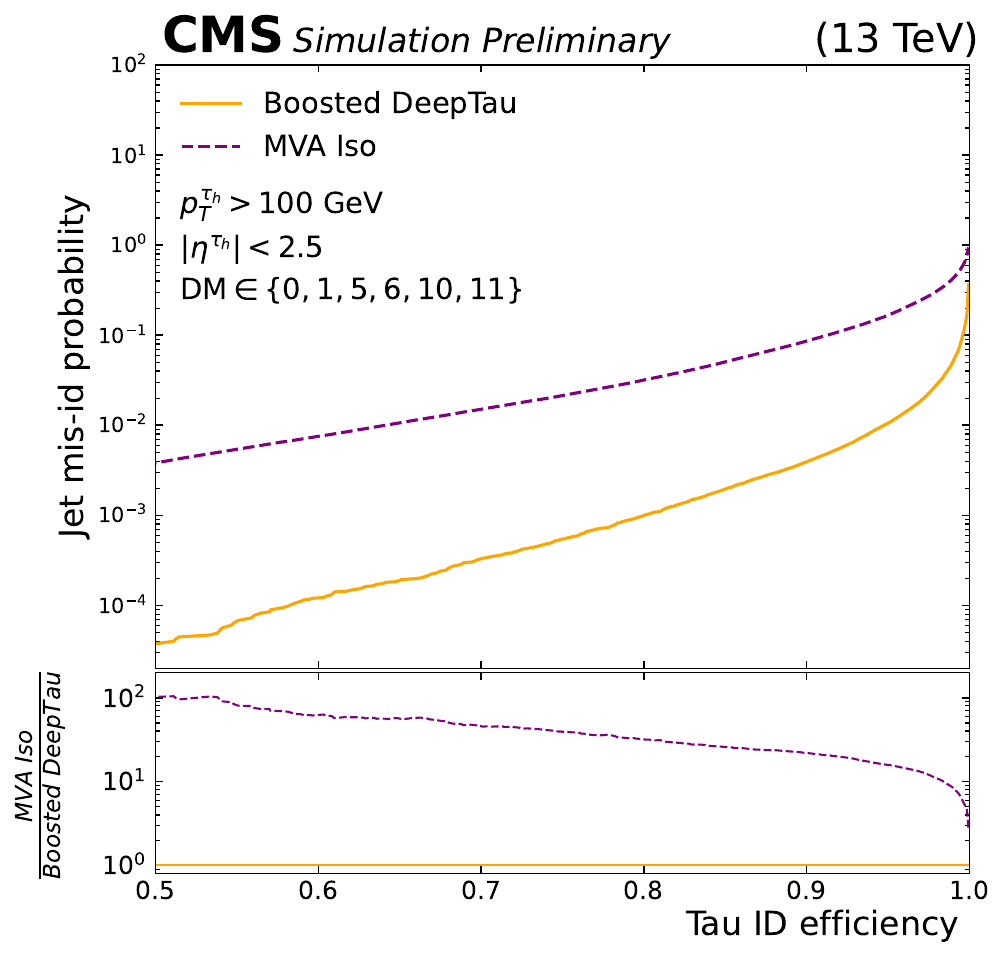}
\caption{\textsc{BoostedDeepTau} shows a discrimination power that is a factor of 2--4 ($>10$) better than the MVA Iso discriminator evaluated on individual reconstructed $\tau_{\mathrm{h}}$ from a boosted ditau system for $p_{\mathrm{T}}^{\tau_{\mathrm{h}}} < 100~\text{GeV}$ ($>100~\text{GeV}$). Here, $p_{\mathrm{T}}^{\tau_{\mathrm{h}}}$ refers to the transverse momentum of the individual reconstructed $\tau_{\mathrm{h}}$ from a boosted ditau system. The Boosted DeepTau ID efficiency is estimated from a spin-0 $\mathrm{X \rightarrow HH \rightarrow bb\tau\tau}$ MC sample (with $M_{X} = 3.5~\text{TeV}$). The jet misidentification probability is estimated from $\mathrm{t}\bar{\mathrm{t}}$+jets MC samples.
\label{fig:roc}
}
\end{figure}

\begin{figure}[tbp]
\centering
    \includegraphics[width=0.48\textwidth]{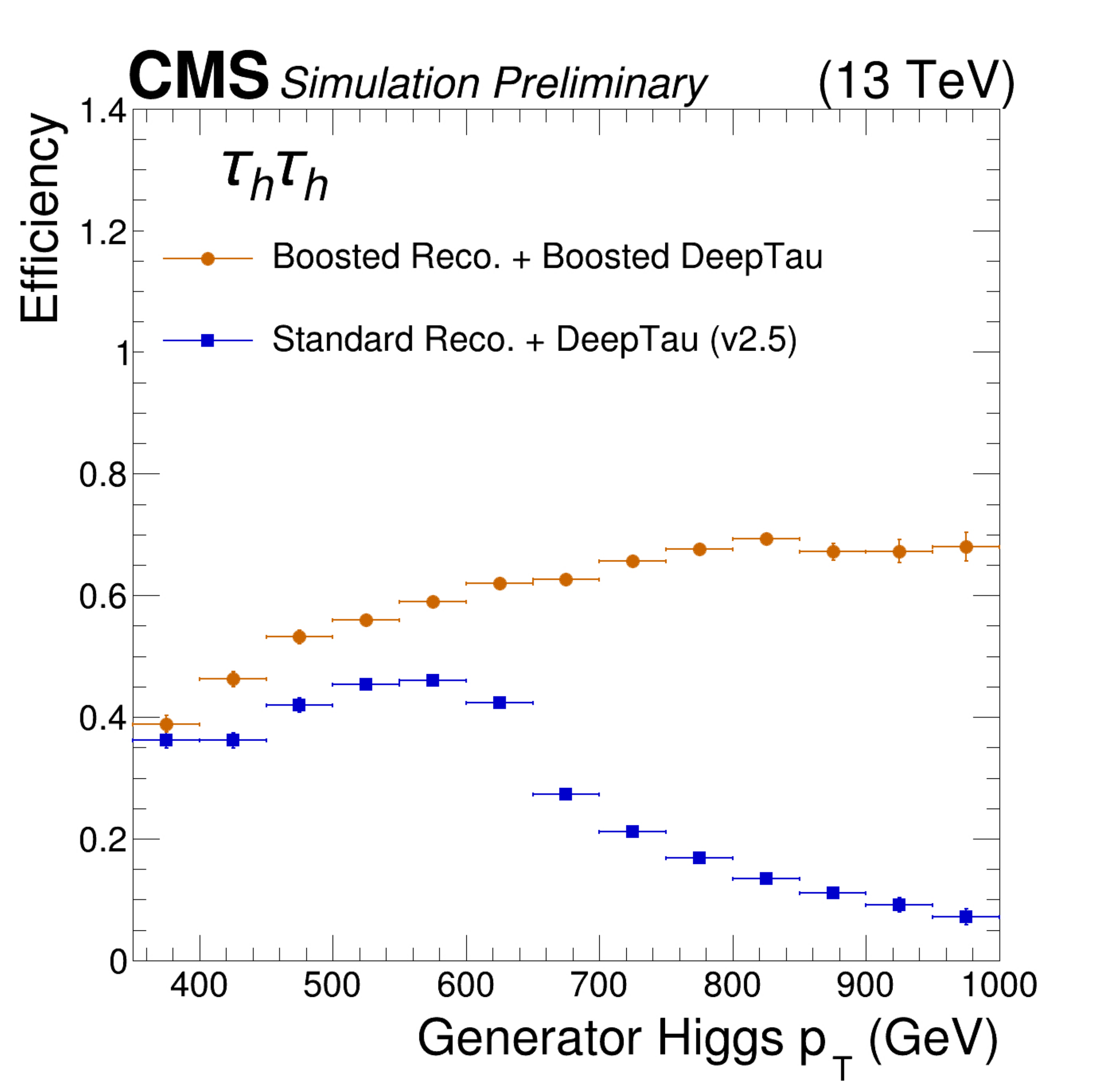}
    \includegraphics[width=0.48\textwidth]{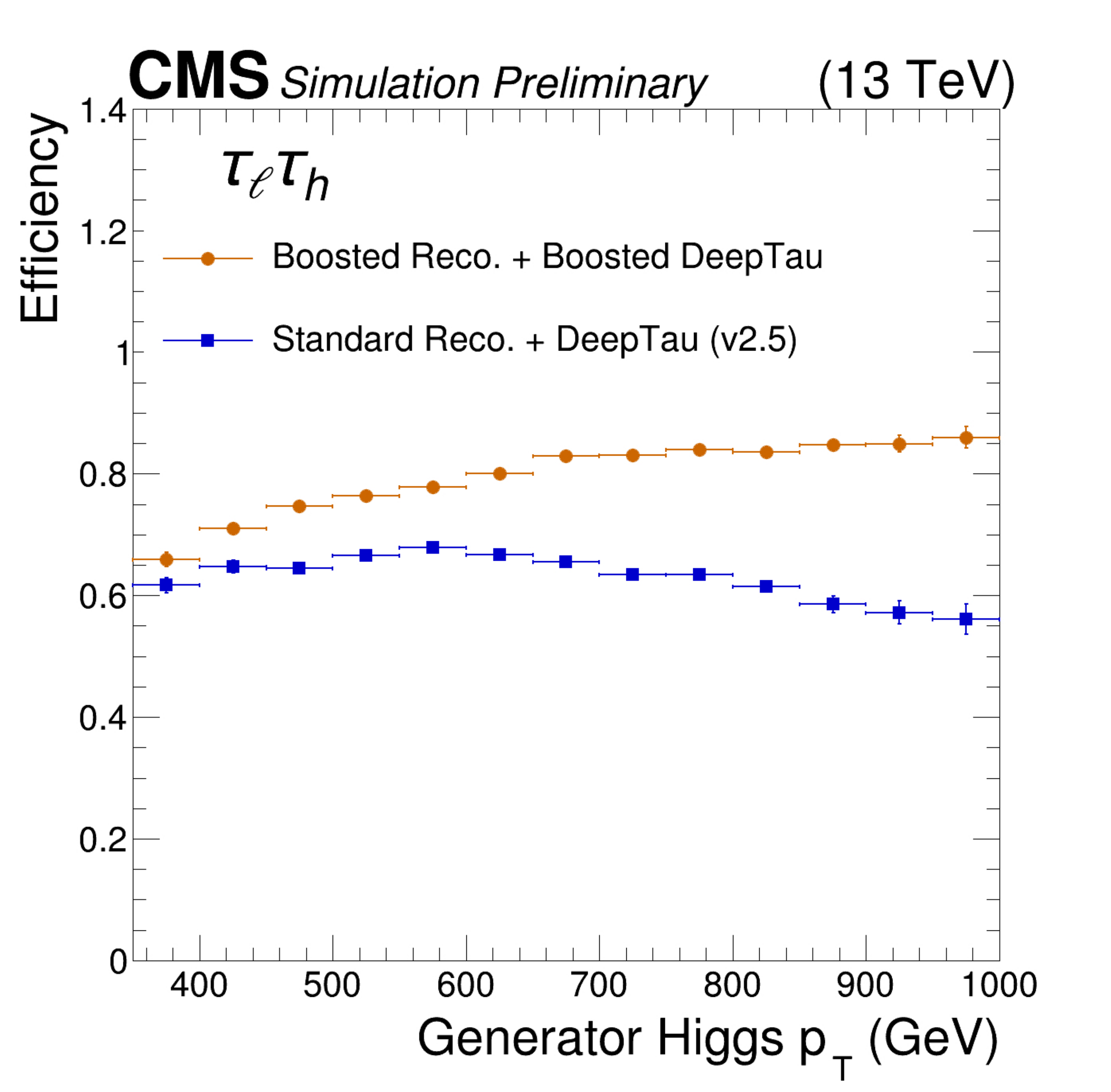}
\caption{H $\rightarrow \tau\tau$ reconstruction and identification efficiency as a function of generator–level Higgs boson $p_{\mathrm{T}}$ is evaluated using spin-0 $\mathrm{X \rightarrow HH \rightarrow bb\tau\tau}$ MC samples (mass of $X = 1.6~\mathrm{TeV}$). The reconstructed $\tau_{\mathrm{h}}$ are required to have $p_{\mathrm{T}} > 20~\mathrm{GeV}$, $|\eta| < 2.5$, and to pass the ``Loose'' working point of the respective identification algorithms, which corresponds to a per-$\tau_{\mathrm{h}}$ signal efficiency exceeding 95\%.
\label{fig:sigeff}
}
\end{figure}

Simillarly, the decay products of the $\tau$-lepton pair,
whether fully hadronic ($\tau_h\tau_h$) or semileptonic ($\ell\tau_h$), also tend to overlap spatially, making di-$\tau$ reconstruction particularly challenging. The reconstruction of such
Lorentz-boosted $\tau$ leptons was first addressed in a similar CMS search using
$8~\mathrm{TeV}$ proton--proton collision data~\cite{CMS:2015fjg}. Higgs bosons
decaying to one or two hadronically decaying $\tau$ leptons are reconstructed
using a dedicated boosted-$\tau$ algorithm that considers subjets of the jets
clustered with the Cambridge--Aachen algorithm and applies a mass-drop criterion
to seed the standard hadron-plus-strips (HPS) procedure. A key element of this
analysis is the new \textsc{BoostedDeepTau}~\cite{CMS-DP-2025-047}
convolutional-neural-network tagger, developed specifically to identify highly
Lorentz-boosted $\tau$ leptons. Built upon the \textsc{DeepTau}~\cite{CMS:2022prd} architecture, it
incorporates both low-level detector information and a set of 42 high-level
variables describing kinematics, isolation, and lifetime features. Trained on
simulated Run~2 samples of signal and major backgrounds,
\textsc{BoostedDeepTau} improves background rejection by a factor 2--4 at
$p_{\mathrm{T}}<100~\mathrm{GeV}$ and by more than an order of magnitude at
higher momenta. Performance summaries are shown in Figs.~\ref{fig:roc}
and~\ref{fig:sigeff}. In addition, this analysis also considers AK4-seeded
$\tau$ candidates identified with the \textsc{DeepTau} algorithm.


The SM processes with the largest contributions to the background are $\mathrm{t\bar{t}}$ (top quark--antiquark production), $\mathrm{W+jets}$, and $\mathrm{Z+jets}$ production, although the precise composition depends on the specific channel under consideration.



\section{ Analysis design}
\label{sec:bkg_estimation_results}
The vector $\,\vec{p}_{\mathrm{T}}^{\mathrm{miss}}$ is defined as the negative of the vector sum of the transverse momenta of all reconstructed particles associated with the primary vertex, and $p_{\mathrm{T}}^{\mathrm{miss}}$ is its magnitude. Events are selected using missing transverse momentum triggers with thresholds above $110~\mathrm{GeV}$, together with an offline requirement of $p_{\mathrm{T}}^{\mathrm{miss}}>180~\mathrm{GeV}$ to ensure high signal efficiency. Each event must contain one Higgs boson candidate decaying to $\ell\tau$ or $\tau\tau$, and a second Higgs boson reconstructed as a large radius jet ($R = 0.8$). The four-momentum of the di-$\tau$ system is reconstructed with the \textsc{FastMTT}~\cite{Matyszkiewicz:2025gal} algorithm, which uses a dynamical-likelihood approach with a collinear approximation. The resonance mass is defined as the invariant mass of this di-$\tau$ system and the Higgs-jet candidate, and only events with $M_{\mathrm{X}}>750~\mathrm{GeV}$ are kept. The overall selection efficiency for simulated spin-0 resonances ranges from 2--14\% in the $\tau\tau$ channel and 4--11\% in the $\ell\tau$ channel.
\begin{figure}[tbp]
\centering
    \includegraphics[width=0.48\textwidth]{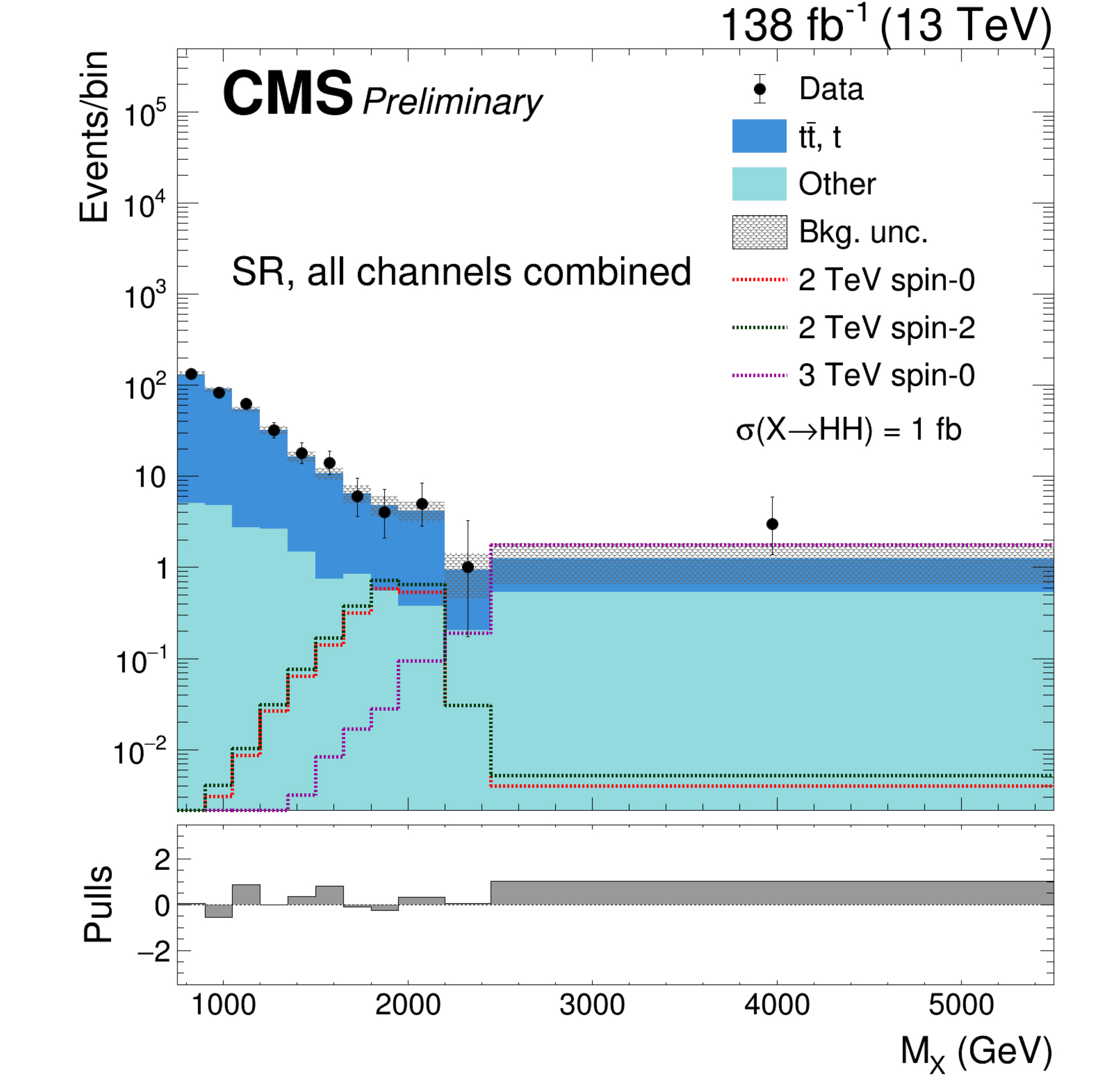}
    \includegraphics[width=0.48\textwidth]{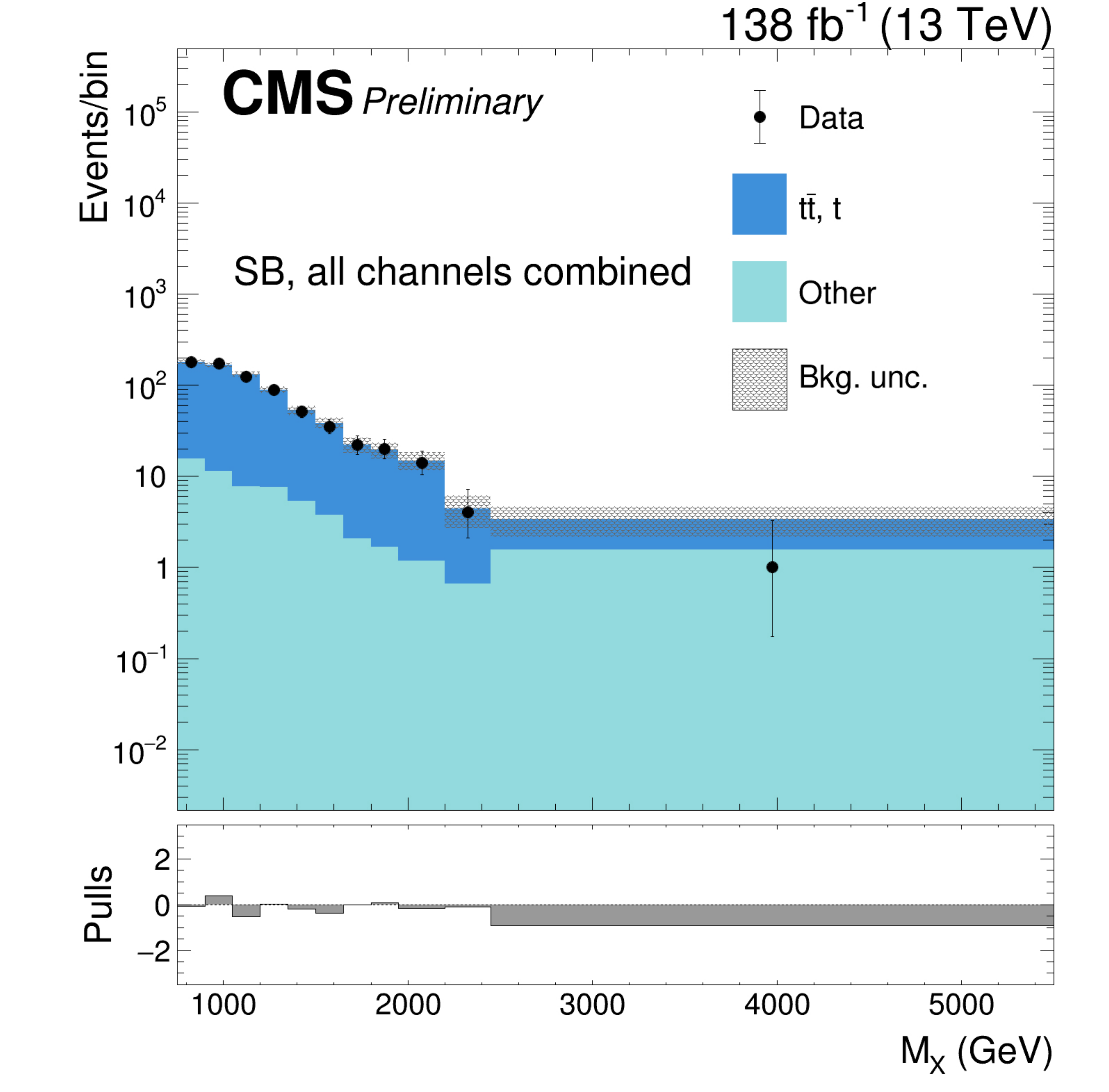}
\caption{Post-fit reconstructed mass distribution of resonance $X$ in the SR (left) and SB (right) after applying all selection criteria, for the sum of the $\tau_{\mathrm{h}}\tau_{\mathrm{h}}$ and $\ell\tau_{\mathrm{h}}$ channels. Minor background contributions are grouped into a single category labeled ``others''. The lower panels show the ``Pull'' defined as $\text{(Data - Prediction)/Total uncertainty}$.
\label{fig:postfit}
}
\end{figure}
Following the full event selection, $\mathrm{t\bar{t}}$ becomes the dominant background in both channels. To improve the modeling of $\mathrm{t\bar{t}}$, an unconstrained bin-by-bin multiplicative parameter is introduced for each bin of the reconstructed resonance mass ($M_{X}$) distribution. These parameters adjust the  $\mathrm{t\bar{t}}$ yield independently in every ($M_{X}$) bin, simultaneously in the SR and the SB, thereby constraining both the normalization and shape of $\mathrm{t\bar{t}}$ using data in the SB for a given channel. A simultaneous binned likelihood fit is performed for signal extraction. Systematic uncertainties affecting normalization or shape of the fit templates are considered, such as those associated with luminosity, trigger and lepton efficiencies, jet energy scale and resolution, pileup modeling, and $\tau$-identification scale factors for both standard and boosted reconstruction. For $\mathrm{t\bar{t}}$, the uncertainty is entirely driven by the data constrained bin-by-bin parameters.

\begin{figure}[tbp]
\centering
    \includegraphics[width=0.48\textwidth]{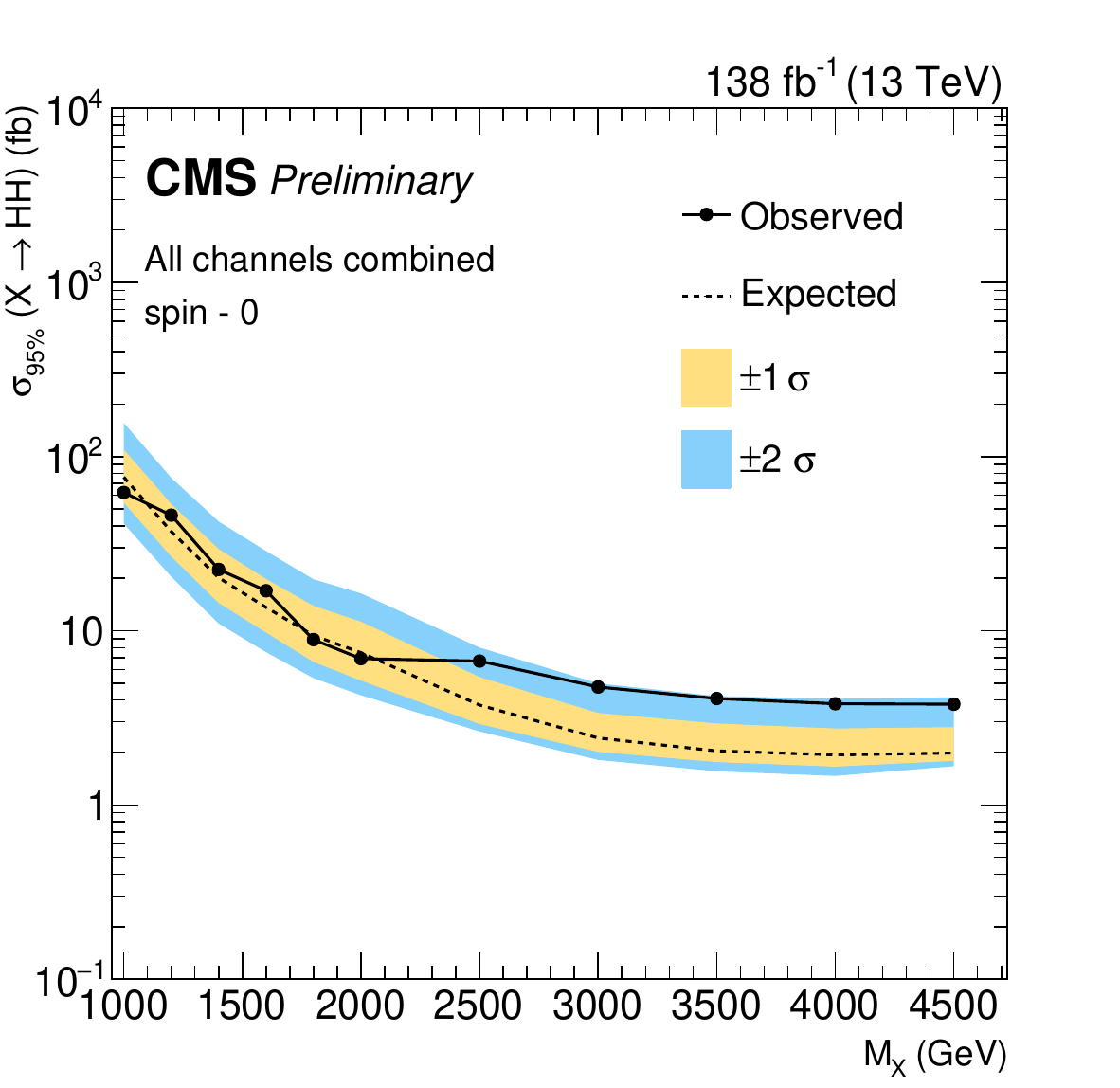}
    \includegraphics[width=0.48\textwidth]{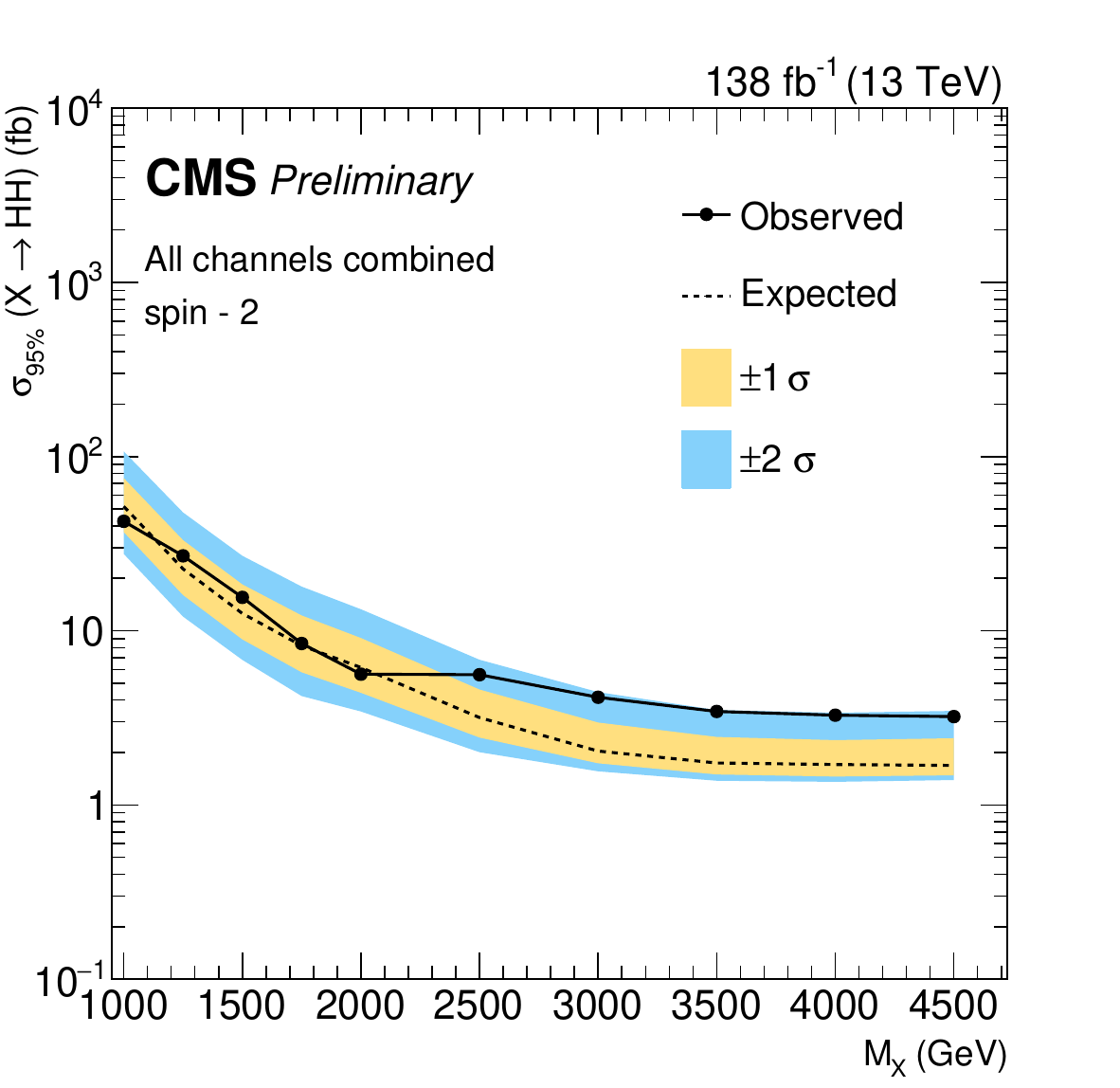}
\caption{Expected and observed upper limits at 95\% CL on the production cross section of resonant $HH$ production for a spin-0 (left) and spin-2 (right) narrow resonance.
\label{fig:limit}
}
\end{figure}

The post-fit distributions in \(M_{X}\) are shown in Fig.~\ref{fig:postfit} for both the SR and SB. 
The data are found to be in agreement with the SM predictions within the total uncertainties. Upper limits are therefore set on the production cross section of a heavy resonance \(X\) decaying to a pair of Higgs bosons independently for spin-0 and spin-2. The plots for this can be seen in Fig.~\ref{fig:limit}. This analysis improves upon the spin-0 full Run~2 low-mass ATLAS search~\cite{aad2020reconstruction} for resonance masses above 1.4~TeV, and the spin-0 full Run-2 boosted ATLAS search~\cite{Run2atlasheavybbtautau} across the entire region considered in that analysis. The results presented here constitute the most sensitive limits to date in the resonance mass range of 1.4 to 4.5~TeV.  
\label{sec:summary}

\section{Summary}

A search for heavy resonant HH pair production in the $\mathrm{b\bar{b}}\tau\tau$ final state is performed using the full Run~2 CMS dataset of 138$\mathrm{fb}^-1$ at $\sqrt{s}=13$ TeV, targeting masses between 1 and 4.5 TeV. Lorentz boosted Higgs decays are reconstructed with advanced machine-learning techniques for merged  $\mathrm{b\bar{b}}$ jets and boosted \(\tau\) leptons. No significant excess is observed, and 95\%\,CL upper limits are set for spin-0 and spin-2 hypotheses. These results are the most sensitive constraints to date on $\mathrm{X}\rightarrow\mathrm{HH}\rightarrow \mathrm{b}\bar{\mathrm{b}}\,\tau^{+}\tau^{-}$ for masses above 1.4~TeV. Additional details of this result can be found in Ref.~\cite{CMS-PAS-B2G-24-014}.


\begin{thebibliography}{14}

\bibitem{ATLAS:2013xga}
G.~Aad et~al. (ATLAS), {Evidence for the spin-0 nature of the Higgs boson using ATLAS data}, Phys. Lett. B \textbf{726}, 120 (2013), \texttt{1307.1432}. \doiwoc{10.1016/j.physletb.2013.08.026}

\bibitem{CMS:2012qbp}
S.~Chatrchyan et~al. (CMS), Observation of a new boson at a mass of {125~GeV} with the {CMS} experiment at the {LHC}, Phys. Lett. B \textbf{716}, 30 (2012), \texttt{1207.7235}. \doiwoc{10.1016/j.physletb.2012.08.021}

\bibitem{randall1999large}
L.~Randall, R.~Sundrum, A large mass hierarchy from a small extra dimension, Phys. Rev. Lett. \textbf{83}, 3370 (1999), \texttt{hep-ph/9905221}. \doiwoc{10.1103/PhysRevLett.83.3370}

\bibitem{goldberger1999modulus}
W.D. Goldberger, M.B. Wise, Modulus stabilization with bulk fields, Phys. Rev. Lett. \textbf{83}, 4922 (1999), \texttt{hep-ph/9907447}. \doiwoc{10.1103/PhysRevLett.83.4922}

\bibitem{davoudiasl2000phenomenology}
H.~Davoudiasl, J.~Hewett, T.~Rizzo, Phenomenology of the {R}andall--{S}undrum gauge hierarchy model, Phys. Rev. Lett. \textbf{84}, 2080 (2000), \texttt{hep-ph/9909255}. \doiwoc{10.1103/PhysRevLett.84.2080}

\bibitem{2016cmsheavybbtautau}
A.M. Sirunyan et~al. (CMS), Search for heavy resonances decaying into two {Higgs} bosons or into a {Higgs} boson and a {W} or {Z} boson in proton-proton collisions at {13~TeV}, JHEP \textbf{01}, 051 (2019), \texttt{1808.01365}. \doiwoc{10.1007/JHEP01(2019)051}

\bibitem{aad2020reconstruction}
G.~Aad et~al. (ATLAS), Reconstruction and identification of boosted di-$\tau$ systems in a search for {H}iggs boson pairs using 13 {T}e{V} proton-proton collision data in {ATLAS}, JHEP \textbf{2020}, 1 (2020), \texttt{2007.14811}. \doiwoc{10.1007/JHEP11(2020)163}

\bibitem{Run2atlasheavybbtautau}
G.~Aad et~al. (ATLAS), Search for resonant and non-resonant {H}iggs boson pair production in the $\text{b}\bar{\text{b}}\,\tau^+\tau^-$ decay channel using 13 {TeV} $\text{pp}$ collision data from the {ATLAS} detector, JHEP \textbf{07}, 040 (2023), \texttt{2209.10910}. \doiwoc{10.1007/JHEP07(2023)040}

\bibitem{qu2020jet}
H.~Qu, L.~Gouskos, Jet tagging via particle clouds, Phys. Rev. D \textbf{101}, 056019 (2020), \texttt{1902.08570}. \doiwoc{10.1103/PhysRevD.101.056019}

\bibitem{CMS:2015fjg}
V.~Khachatryan et~al. (CMS), Search for narrow high-mass resonances in proton-proton collisions at $\sqrt{s} = 8~\text{TeV}$ decaying to $z$ and {H}iggs bosons, Phys. Lett. B \textbf{748}, 255 (2015), \texttt{1502.04994}. \doiwoc{10.1016/j.physletb.2015.07.011}

\bibitem{CMS-DP-2025-047}
{CMS Detector Performance Summary} CMS-DP-2025-047, CERN (2025), \urlstyle{tt}\url{https://cds.cern.ch/record/2941434}

\bibitem{CMS:2022prd}
A.~Tumasyan et~al. (CMS), Identification of hadronic tau lepton decays using a deep neural network, JINST \textbf{17}, P07023 (2022), \texttt{2201.08458}. \doiwoc{10.1088/1748-0221/17/07/P07023}

\bibitem{Matyszkiewicz:2025gal}
W.~Matyszkiewicz, A.~Kalinowski, {Tau-pair invariant mass estimation using maximum likelihood estimation and collinear approximation}, Acta Phys. Polon. Supp. \textbf{18}, 5 (2025). \doiwoc{10.5506/APhysPolBSupp.18.5-A21}

\bibitem{CMS-PAS-B2G-24-014}
{CMS Physics Analysis Summary} CMS-PAS-B2G-24-014, CERN (2025), \urlstyle{tt}\url{https://cds.cern.ch/record/2941053}

\end{thebibliography}

\end{document}